\documentstyle[12pt,aasms4,epsf]{article}

\def\sun{\hbox{$\odot$}}

\def\lae{\mathrel{\hbox{\rlap{\hbox{\lower4pt\hbox{$\sim$}}}\hbox{$<$}}}}
\def\gae{\mathrel{\hbox{\rlap{\hbox{\lower4pt\hbox{$\sim$}}}\hbox{$>$}}}}

\def\min{\hbox{$^\prime$}}

\def\etal{{\it et al. }}

\begin{document}

\title{THE OLD-HALO GLOBULAR CLUSTER SYSTEM IN NGC 1275}
\author{Denise Kaisler and William E. Harris\altaffilmark{1}}
\affil{Department of Physics and Astronomy, McMaster University \\
Hamilton, Ontario, L8S 4M1, Canada\\
{\it kaislerd@physics.mcmaster.ca, harris@physics.mcmaster.ca}}
\author{Dennis R. Crabtree\altaffilmark{1}}
\affil{Dominion Astrophysical Observatory, Herzberg Institute of Astrophysics \\
5071 West Sannich Road, Victoria, British Columbia \\
V8X 4M6, Canada\\
{\it crabtree@dao.nrc.ca}}
\author{Harvey B. Richer}
\affil{University of British Columbia, Department of Geophysics and Astronomy\\
Vancouver, British Columbia, V6T 1Z4, Canada\\
{\it richer@astro.ubc.ca}}

\altaffiltext{1}{Visiting Astronomer, Canada-France-Hawaii Telescope, operated by the National Research Council of Canada, le Centre national de la Recherche Scientifique de France, and the University of Hawaii.}
 
\begin{abstract}
We present the results of a deep photometric study of the outer halo of 
NGC 1275, the highly active cD galaxy at the center of the Perseus cluster. 
We find a modest excess of faint ($R > 22.5$) starlike objects in its halo,
indicating a population of old-halo globular clusters. However, the total estimated cluster population corresponds to a specific frequency of
$S_N = 4.4 \pm 1.2$, no larger than that of normal giant ellipticals and three times lower than that of other central cD galaxies such as M87.  We discuss several ideas for the origin of this galaxy. Our results reinforce the view that high $S_N$ (ie: highly efficient globular cluster formation) is not associated with cooling flows, or with recent starburst or merger phenomena.  
\end{abstract}
\clearpage

\section{Introduction}

Galaxies are known to exhibit wide differences in the relative numbers
of their globular clusters (GCs).  A useful quantitative measure of this effect
is the specific frequency (\markcite{har81}Harris \& van den Bergh 1981; \markcite{har91}Harris 1991), 
\begin{equation}
S_N = N_t \cdot 10^{0.4(15+M_V^T)}
\label{specific}
\end{equation}
defined as the ratio of total cluster population $N_t$  to galaxy luminosity $M_V^T$. $S_N$ represents the overall efficiency of cluster formation (or
accretion) averaged over the galaxy's history. Spiral galaxies tend to have the smallest populations ($S_N \sim 1-3$), while normal ellipticals rank somewhat higher ($S_N \sim 4-6$; see \markcite{har91}Harris 1991). In this respect, an outstandingly interesting group consists of the cD galaxies  found in the centers of rich clusters. Although some of these have $S_N$ ratios comparable to ordinary ellipticals, others have ratios as high as $S_N \sim 15$. The suggested origins for these very large GC populations have included formation out of cooling flows (\markcite{fab84}Fabian, Nulsen, \& Canizares 1984), mergers of spirals (\markcite{sch87}Schweizer 1987; \markcite{ash92}Ashman \& Zepf 1992; \markcite{zep93}Zepf \& Ashman 1993), accretion by tidal stripping and cannibalism of neighboring galaxies (e.g. \markcite{muz88}Muzzio 1988), or a high GC formation efficiency during the protogalactic era (\markcite{har91}Harris 1991). Each of these options has distinct difficulties in explaining the data: a thorough summary of the issues for these cD galaxies is given by \markcite{hpm95}Harris, Pritchet, \& McClure (1995 = HPM95). 

NGC 1275 (Perseus A) potentially provides an interesting touchstone for many
of the arguments surrounding the issue of GC formation, since it possesses a
combination of properties that are rarely found even among central giant ellipticals. It is an extremely luminous cD-type galaxy at the center of
the rich Perseus cluster (Abell 426), it has one of the largest
known cooling flows ($\sim$ 200 M\sun/yr; \markcite{fab81}Fabian \etal 1981), a highly active
Seyfert-type nucleus, and has undergone very obvious GC formation in its central regions in the past $\sim 10^9$ y (\markcite{hol92}Holtzman \etal 1992; \markcite{rich93}Richer \etal 1993). 
\markcite{zep95}Zepf \etal (1995) note the unusual distended shape of this 
Seyfert galaxy and suggest recent merger events as the cause.
Many other features add to the interest of this galaxy. A system of luminous filaments (see Fig. \ref{galaxy}) stretches more than 20 kpc from the galaxy's center. These filaments have a systemic radial velocity of $\sim$ 5300 km s$^{-1}$, the same as the main body of NGC 1275 (the `low velocity' or LV system).They may have been born out of the cooling flow and ionized 
either by nonthermal emission from the nucleus (\markcite{ken79}Kent \& Sargent 1979) 
or by shocks within the LV material itself (see \markcite{he89}Heckman \etal 1989 for a review). The nucleus is a region of high H$\alpha$ emission and is host to both molecular (\markcite{laz89}Lazareff \etal 1989) and neutral (\markcite{jaf90}Jaffe 1990) hydrogen.  

Equally important is the existence of a `high velocity' (HV) system of giant HII clouds with V$_o$ $\sim$ 8300 km s$^{-1}$ which occupies the N and NW regions of NGC 1275. Various theories exist regarding the relationship of the LV and HV components. The initial supposition (\markcite{baa54}Baade \& Minkowski 1954) was that the HV system represents the gaseous remains of another galaxy, perhaps a large spiral, merging with the underlying giant elliptical. Alternately, 
\markcite{dey73}De Young, Roberts, \& Saslaw (1973) proposed that the two components 
constitute a chance superposition of two galaxies that do not interact. 
Still another proposal (\markcite{bur65}Burbidge \& Burbidge 1965) is that the HV system 
was ejected from the core of NGC 1275.
  
Later evidence has served to place the HV system at least partially in the foreground: HI associated with the HV system has been found in absorption 
in the spectrum of the nuclear radio source 3C 84 (\markcite{dey73}De Young, Roberts, \& Saslaw 1973), and IUE spectra show HV-related Ly$\alpha$ in front of the nucleus 
(\markcite{rub77}Rubin \etal 1977; \markcite{bri82}Briggs, Snijders, \& Boksenberg 1982). Similarly, absorption of NGC 1275 starlight has been associated with the areas of strongest HV emission (\markcite{nor93}N\o rgaard-Nielsen \etal 1993). This evidence, combined with the lack of a symmetrical counterjet has made the ejection hypothesis unlikely. 
New evidence for the interacting-galaxy hypothesis is provided by 
Boroson's (\markcite{bor90}1990) discovery of stellar absorption in the Ca II $\lambda$8948 line
of the HV system, together with a new velocity curve that suggests 
a velocity gradient across the HV gas.  

The recent discovery of several compact blue objects in the galaxy core (\markcite{hol92}Holtzman \etal 1992) has attracted much attention. The initial identification of many of these objects as young ($<$ 1 Gyr) globular clusters has been substantiated by later photometry and spectroscopy (\markcite{rich93}Richer \etal 1993; \markcite{zep95}Zepf \etal 1995). Condensation out of the cooling flow, past mergers, and encounters with the HV system have all been suggested as possible causes for their origin.  Holtzman \etal (\markcite{hol92}1992) and \markcite{zep95}Zepf \etal (1995) contend that the young GCs are
evidence of a merger with another galaxy, which would have brought in a fresh supply of relatively cool gas. It is not obvious whether or not the HV system represents part of the merger remnant, although \markcite{righ93}Richer \etal (1993) point out that the high relative velocity of the two systems ($\sim$ 3000 km/s) gives
an interaction timescale of only $\sim$ 10$^6$ yr, which does not agree with the proposed cluster ages of 0.5 Gyr (\markcite{zep95}Zepf \etal 1995). In addition, examination of the velocity curve for the HV system suggests that it is only just beginning to
come in contact with the main body of NGC 1275 (\markcite{bor89}Boroson 1990).
Lester \etal (\markcite{les95}1995) identify thermal emission from a dust component
and claim that its presence argues in favor of a recent merger.
Thus, the nuclear activity may be the visible result of a merger that took
place in the last $\sim 1$ Gyr and is by now nearly complete, with the HV system
being a still more recent arrival.
 
{\it If} globular clusters form efficiently out of spiral mergers (e.g. \markcite{zep93}Zepf \& Ashman 1993) or cooling flows (\markcite{fab84}Fabian \etal 1984), and if the very rich cluster systems in cD galaxies built up this way, then we might anticipate that the `old', outer halo of NGC 1275 would contain a populous GC system. An additional reason for expecting NGC 1275 {\it a priori} to have a large cluster population is the strong correlation between high $S_N$ and environment. Of the half-dozen known high-$S_N$ systems (\markcite{hpm95}HPM95; \markcite{cot95}C\^ot\'e 1995), all are {\it very} near the dynamical centers of {\it rich} galaxy clusters. Specifically, these are the cD galaxies in Virgo, Fornax, Hydra I, Coma, A2052, and A2107. NGC 1275 is an extremely luminous galaxy quite near the dynamical center of the Perseus cluster: at a redshift $V_0 = 5316 \pm 78$ km/s, its velocity is little different from the mean cluster redshift of $5451 \pm 168$ km/s (\markcite{bir94}Bird 1994). To compare this peculiar velocity to that of similar cD galaxies, Bird uses Gebhardt and Beers' (\markcite{geb91}1991) Z score parameter  
\begin{equation}
Z_{GB} = {v_{cD} - C_{BI}\over S_{BI} },
\end{equation}
where C$_{BI}$ and S$_{BI}$ are the biweight estimators 
of mean cluster velocity and velocity dispersion
(these are more robust than the usual Gaussian centroid and dispersion,
which tend to fail when applied to non-Gaussian distributions).
Bird lists the Z-score for NGC 1275 as $Z=-0.108$ (corrected for substructure),
i.e. roughly a tenth of a standard deviation from the Perseus centroid. 
When we add the fact that NGC 1275 is at the center of an
extensive X-ray halo, the evidence is strong that it is sitting very nearly
at the center of Perseus.  By association with the other cD galaxies mentioned
above, its old-halo cluster population should then be very rich.
If, on the other hand, its specific frequency is low, then 
several other models for GC formation in such galaxies might be viable.

There is little information available on the old-halo clusters in NGC 1275: 
to date, the sole study addressing the old GC population is that of
\markcite{nor93}N\o rgaard-Nielsen \etal (1993).  From images of the inner 1\min ~of the galaxy,
they suggest by extrapolation that NGC 
1275 may have a total GC population 2/3 as great as that 
of M87 (N$_t$ $\sim$ 10$^4$), in which case it would have a normal $S_N$.
However, the internal uncertainties in their data are too great to permit 
definitive conclusions.
 
In this paper, we present the results of a deeper and more extensive survey of the outer halo of NGC 1275, and discuss its implications for GC formation scenarios.
 
\section{Observations and Data Reduction}

The images for this study were obtained 1993 September 21 at the prime focus 
of the Canada-France-Hawaii Telescope (CFHT), with the High Resolution Camera (HRCam; \markcite{mcc89}McClure \etal~1989) and the Loral3 CCD binned $2 \times 2$ to give an image scale 0.216 arcsec pixel$^{-1}$. The usable photometric field of the camera was $750 \times 750$ binned pixels,or 2.7 arcmin on a side. A $20 \times 300$ sec exposure series in $R$ nearly centered on NGC 1275 was obtained as the primary material for our study (attempts to obtain comparably deep images in other filters, as well as a background control field, were unfortunately terminated by weather and instrument problems). The $R$ band was chosen because it is near the peak sensitivity of the CCD, and is well matched to the moderately red colors of our target objects (the old globular clusters).
The series of images was re-registered and median-combined to yield the composite picture shown in Fig.~\ref{galaxy}; the mean FWHM (seeing) on this combined frame is 3.4 px (0.75 arcsec).

Photometric calibration was carried out by transferring short exposures of
Landolt (\markcite{lan92}1992) standard stars onto our individual NGC 1275 exposures.  
The resulting solution for the standard stars was 
\begin{equation}
R = r - 0.10 X - 0.013 (V-R) + const
\label{calibrate}
\end{equation}
where $r$ is the instrumental magnitude through a large (15 px) aperture and
$X$ is the airmass.  This relation reproduced 
the standard-star values to within $\pm0.01$ mag.
Since the color coefficient is very small and we have only one bandpass for our
program field, we neglected the $V-R$ term.
Final PSF-fitting photometry of all the objects on the composite frame
was carried out only after removal of the overall contours of the
galaxy light. A model for the isophotal contours was created using
the STSDAS routines isophote.ellipse and isophote.bmodel. This model
was then subtracted from the picture to eliminate 
the major part of the light gradient from the body of NGC 1275.
A sequence similar to that described by \markcite{fis90}Fischer \etal (1990) was then
implemented, whereby passes of DAOPHOT/ALLSTAR (\markcite{ste92}Stetson 1992; \markcite{ste90}Stetson, Davis,
\& Crabtree 1990) 
were alternated with smoothing and removal of the remaining background light
with the fast ring-median filter of Secker (\markcite{sec95}1995). 
Two iterations of this sequence proved to be sufficient 
to locate virtually all of the starlike objects in the field. 
  
\section{Analysis}

To evaluate the photometric completeness as a
function of magnitude and position on the frame and to help set parameters for
rejection of nonstellar images, we carried out a series of simulations. 
First, all the detected objects and the galaxy light were subtracted 
to produce a frame bare of any features except background sky noise. 
With this blank frame we then recreated a dozen simulated frames, 
adding a total of 3600 stars in groups of 300 over a wide range of magnitudes. 
These were spatially distributed 
to mimic the light profile of the galaxy (see below), 
following a simple power-law distribution centered on the main galaxy. 
The results of these experiments for the completeness fraction $f$
(the fraction of objects recovered at any magnitude)
are summarized in Figure \ref{compfunc} and Table 1. 
Fig.~\ref{compfunc} displays the experimental results, fitted by Pritchet's
interpolation function (see \markcite{fle95}Fleming \etal 1995),
\begin{equation}
f = {1\over{2}}\left[1- {\alpha (R - R_{lim})\over{\sqrt{1 + \alpha^{2}(R - R_{lim})^{2}}}}\right]
\end{equation}
The limiting magnitude (i.e. the level at which $f$ drops to 50\%)
is $R \simeq 24.43$, a value nearly independent of radial 
location for our main region of
interest ($r > 100$ px, outside the central region where the galaxy light dominates;
see below).  All fainter objects were discarded from our lists.  Note that the
completeness function drops quite steeply, so that for $R < 24$ the photometry
is virtually 100\% complete.
 
For any galaxy at the distance of NGC 1275, globular clusters appear as faint
starlike objects sprinkled around its halo. 
Yet due to the presence of the optical filaments, the NGC 1275 field is
photometrically much more complex than that of a more normal giant elliptical.
These features caused the DAOPHOT algorithms to fit PSFs to the 
clumpier regions of the filaments, thus generating numerous spurious detections. When these are added to other types of contaminating objects on 
the frame (faint background galaxies, Perseus dwarfs, etc.), 
the majority of the detected objects in the raw lists turn out to 
be nonstellar and need to be eliminated as completely as possible. 
To accomplish this, we used the radial moments $r_1$ and $r_{-2}$, 
which measure image wing spread and central concentration respectively, 
along with an isophotal magnitude $R_i$  (\markcite{hap91}Harris \etal 1991).
These plots are shown in Figure \ref{moments}, where the starlike objects lie along narrow sequences. The three parameters are so well correlated in this field that they are largely redundant, and the criterion we adopted was simply to reject objects with $r_1$ $>$ 2.5, since this moment gave the clearest empirical discrimination of nonstarlike objects. This step eliminated most of the extended objects and multiple detections from the list. Finally, the remaining 349 objects (candidate GCs) were then examined one by one against the original image. All objects within 5 pixels of the frame edge, or which were associated with bad pixels, filaments, or were otherwise anomalous to careful 
inspection, were also rejected.  This culling procedure left 294 final candidates.

Since the old-halo globular clusters have such a similar 
luminosity function (GCLF) among all types of large galaxies (especially
among the giant ellipticals; e.g., \markcite{har91}Harris 1991; \markcite{hap91}Harris \etal 1991; \markcite{ajh94} Ajhar \etal
1994; \markcite{whit95}Whitmore \etal 1995), we can make an excellent preliminary guess at the
magnitude range over which we expect to detect a cluster population around NGC
1275.  For the well known Virgo ellipticals the 
GCLF peaks at $R \simeq 23.2 \pm 0.2$ (or $M_R = -7.8$ for $(m-M)_V = 31.0$) and
with a dispersion near $\sigma = 1.4 \pm 0.1$ (\markcite{hap91}Harris \etal 1991; \markcite{whit95}Whitmore \etal 1995; \markcite{har96}Harris, Harris, \& McLaughlin 1996 \markcite{sec93}Secker \& Harris, 1993).  
At a redshift $V_0 \sim 5400$ km/s, NGC 1275 is four times 
farther away; after addition of foreground absorption ($E_{B-V} = 0.15$;
\markcite{bur84}Burstein \& Heiles 1984), we would then expect 
the GCLF turnover to lie at $R \simeq 26.5$,
more than two magnitudes fainter than our photometric limit.  
The GCLF should then start appearing only for $R > 22.5$ and should  
rise rapidly faintward of that level.  Our directly observed result 
(see below) is fully in line with this expected range.

Figure \ref{xypos} shows the positions of the final 294 GC 
candidates over the target magnitude range
$22 < R < 24.43$.  Several gaps in the spatial distribution 
correspond with the positions of the 
bright foreground stars, other large Perseus galaxies, 
and the HV region just northwest of the galaxy core. The latter has been 
associated with regions of high absorption (\markcite{nor93}N\o rgaard-Nielsen \etal 1993). 

The elliptical annuli in Fig.~\ref{xypos} were used to calculate the 
radial distribution, excluding the inner two annuli, for which the number statistics were poor. This distribution is listed in Table 2 and displayed in
Figure \ref{sigma3}, where the mean
number density $\sigma$ of the detected candidates (corrected for areal and
magnitude incompleteness) is plotted against radius.
A modest falloff in $\sigma$ toward larger radii is evident, suggesting that
we have successfully detected 
the presence of an old-halo globular cluster system.
However, the fact that this system was {\it not} obvious from casual inspection of the
images (or indeed from inspection of the `cleaned' distribution in 
Fig.~\ref{xypos}), already suggests that NGC 1275 is not a high-specific-frequency
galaxy like M87.  This suspicion will be borne out in the calculation of $S_N$ to be
made below.

Next we derive the luminosity function. The measured GCLF is shown in Figure \ref{lumfunc}. Here, the outermost three annuli are again used to set the background LF, shown by the dashed line. As expected, the residual population $\phi(R)$ (number of clusters per 0.2-magnitude bin) is negligible for $R < 22.5$ and rises steadily and steeply to fainter magnitudes as is typical for a normal old-halo GCLF. Although the faintest bin shows a rather sudden decrease in $\phi$, this does not represent the turnover point, which would be $\sim 2$ mag fainter. In this last bin, the completeness fraction is near $f \simeq 0.5$ and is clearly the least certain point, although it is still
puzzlingly low. Adopting the turnover level and standard deviation of the GCLF with uncertainties as described above, we find that the observed population brighter than $R$(lim) = 24.4 represents only the top (3.75 $\pm$ 1.00)\% of the total GCLF population.

Since we had no background comparison field, a preliminary estimation of the
background level was obtained simply by taking the total number of objects in the outermost three  annuli and dividing by the total area of these, giving $\sigma_b = 39 \pm 11$ arcmin$^{-2}$.
An alternate approach (e.g. \markcite{har86}Harris 1986) is to fit 
a simple power law 
\begin{equation}
\sigma = \beta R^{\alpha} + \sigma_b
\label{power}
\end{equation}
to the data points and to solve for as many of the parameters as possible directly from the fit.  Since the radial region under scrutiny is not extensive (0.5\min $< r <$ 2.0\min), the data do not constrain the exponent $\alpha$ in any useful way. Instead, we set $\alpha$ by using the relation (\markcite{har93}Harris 1993) between $\alpha$ and the integrated absolute magnitude of the galaxy $M_V^t$:
\begin{equation}
\alpha = (-0.29 \pm 0.03)M_V^t - 8.00.
\label{alpharel}
\end{equation}
[NB: The zero point of this relation appears incorrectly in \markcite{har93}Harris (1993). Here we quote the corrected version.]

The integrated absolute magnitude of NGC 1275, with $V_t$ = 11.59 (\markcite{dev76}de Vaucouleurs, de Vaucouleurs, \& Corwin 1976), is $M_V^t = -23.3$ for a distance modulus $(m-M)_V = 34.9 $ ($H_0 = 75$ km s$^{-1}$ Mpc$^{-1}$ and $A_V=0.6$). However, the spectrum of NGC 1275 is strongly affected by blue light from starburst activity in the core, rendering the galaxy brighter than a similar gE in which no star formation was occurring. \markcite{rom87}Romanishin (1987) concludes that the ratio of excess blue light to that of the old halo is 15\%. Correcting for this, we will adopt $M_V^t(corr)= -23.14$, which produces $\alpha \simeq -1.3$.

Since the scatter around the mean relation is $\pm 0.2$ \markcite{har93}(Harris 1993), we then fit equation (\ref{alpharel}) using $\alpha = (-1.1, -1.3, -1.5)$. Assuming $\alpha$, we then solve for $\sigma_b$ and $\beta$ by straightforward least squares. In what follows, we use unweighted solutions. Weighted solutions would slightly lower $\sigma_b$ and leave the effect of increasing the residual cluster populations by about 20\%. 

Fig.~\ref{sigma3} shows all three solutions and the radial profile. The value $\sigma_b \sim 33$ arcmin$^{-2}$ (for $\alpha = -1.3$) may still be an overestimate of the true background level (2\min ~corresponds to a linear distance of $\sim 40$ kpc, and the halo GCS can extend detectably beyond that radius for particularly large systems, such as M87). However, lacking a direct determination from an adjacent control field, we cannot isolate $\sigma_b$ any more narrowly. Statistical estimates of $\sigma_b$ gathered by modelling starcounts and background galaxies are not of any additional help because of the unusual complexity of the Perseus field in the neighbourhood of NGC 1275.

Next are the corrections to the total population due to area. To obtain the total cluster population over all radii we integrate relation \ref{power}
from $0.5'$ out to a generous limit $5'$ (corresponding to 100 kpc), multiply by the ellipticity area factor $q = 0.8$, and subtract the background value. This calculation yields the values found under $N_{int}$ of Table 3. A rough estimate of the population in the innermost region $r < 0.5'$ may be found by taking a constant (background-subtracted) value of $\sigma \simeq 40$ arcmin$^{-2}$, which gives an additional $\sim 30$ clusters brighter than our limiting magnitude (we emphasize again that this inner population is the {\it old-halo} GCS, and does not include any of the blue objects discovered in the HST survey). Adding these 30 clusters and dividing by ($0.0375 \pm 0.010$) to account for the portion of the GLCF which is fainter than our limiting magnitude, we then derive the total old-halo cluster populations given in Table 3.  

For the specific frequency we then obtain S$_N$ = 4.3 $\pm$ 1.4 -- within the range of the Virgo and Fornax ellipticals (\markcite{har91}Harris 1991), but more than 3 times smaller than the `high-$S_N$' cD systems of which M87 is the prototype. Perhaps a more dramatic illustration of this conclusion is shown in Figure \ref{sigmam87}, in which we show the observed radial profile of the M87 GCS as it would appear at the distance of NGC 1275, normalized to the same total luminosity. If, as noted above, we have significantly overestimated the background level $\sigma_b$, the results for the total population would change, but not dramatically so. Even our shallowest adopted profile ($\alpha = -1.1$) excludes NGC 1275 from the high-$S_N$ ellipticals.  Unless deeper and wider-field data reveal major surprises, it seems clear that as giant ellipticals go, NGC 1275 has a normal or even subnormal specific frequency.
 
\section{Discussion}

All previous observations have shown that the conditions necessary 
to form a high-$S_N$ galaxy are dominant size and central location
within a rich cluster (\markcite{hpm95}HPM95). On these grounds, we might reasonably expect
NGC 1275 to be such a candidate. 
In addition, \markcite{mcl94}McLaughlin, Harris, \& Hanes (1994) conclude that there is a 
moderate correlation between Bautz-Morgan class and S$_N$. 
The BM class of Perseus (II-III) would then also suggest that its central
cD should be well supplied with GCs.  Furthermore, NGC 1275 is undergoing highly active phases of gas accretion and star formation. 
Despite all these conditions, its specific frequency is unexceptional.  This
discrepancy casts some doubt on the view (\markcite{ash92}Ashman \& Zepf 1992, \markcite{zep93}Zepf \& Ashman 1993, \markcite{zep95}Zepf \etal 1995) that merger or starburst systems are particularly favorable places for GC formation.

However, our result strongly reinforces the conclusion
of \markcite{hpm95}HPM95 that cooling flows do not produce large numbers of globular clusters in cD galaxies. In this one respect, NGC 1275 is similar to certain other cD systems that they discuss. By inference, it would then seem more likely that its very extensive nuclear activity and young cluster formation are due to accretion or infall of gas from another source.

To explain our result for NGC 1275, we are left with a variety of options:

{\it (1) NGC 1275 was a normal galaxy that has undergone several mergers.}
In this interpretation, we would assume that 
NGC 1275 started as an unexceptional galaxy
(either spiral or elliptical) which by chance was close to the dynamical
center of A 426.  Since then, it has merged with
numerous other galaxies, producing an overall cD-like and highly
luminous product in which the traces of the original galaxy have been all
but erased.  The merger or accretion currently taking place is then only the
latest in a series. 

Spiral galaxies have low specific frequencies ($S_N \simeq 1-3$;\markcite{har91} Harris 1991) and merger products of spirals are likely to be similarly endowed. It is unquestionably true that globular clusters do form in such mergers (\markcite{sch87}Schweizer 1987; \markcite{ash92}Ashman \& Zepf 1992; \markcite{whi95}Whitmore \& Schweizer 1995). But this, by itself, will not necessarily increase the specific frequency of the product galaxy.  The reason is essentially that  $S_N$ is a population ratio of clusters to field stars; therefore, any increase in S$_N$ requires the {\it preferential}
formation of GCs over field stars (which also form during the merger
in large numbers of unbound clumps and associations).  Thus, the final $S_N$ depends critically on how efficiently the merger converts the incoming gas into clusters, as well as on the specific frequencies of the incoming pre-merger galaxies. The cluster formation efficiency is likely to depend on the detailed geometry of the merger, the relative velocities and orientations of the incoming galaxies, the amounts of gas within each, and other factors. 

Averaged over many encounters, it may be reasonable to expect a resulting $S_N$ in the range we observe for NGC 1275 only if the formation of GCs was remarkably efficient and extremely large amounts of gas were brought in by the mergers (\markcite{har95}Harris 1995). As yet, no models exist to calculate these efficiencies. Although it may not be ruled out, we regard this option as an unlikely solution.

{\it (2) NGC 1275 was a high-$S_N$ giant elliptical galaxy that has undergone many accretions.}
This idea is essentially a more conservative version of the previous one.
Here, the initial gE might have had a normal specific frequency
to go along with its centrally dominant position. But since any accreted galaxies
(especially spirals) would have had lower specific frequencies, the eventual
product cD would have been left with a more modest cluster population and
an $S_N$ value that has been diluted down through the mergers by a factor of 2 or more. This idea simply would permit a much lower GC formation efficiency.
Again, the merger we are seeing today would be only the latest accretion event in an ongoing series.

While both of these two scenarios are qualitatively 
consistent with the observations, a few simple calculations cast some 
doubt on them.  If the product galaxy resulted almost totally from spiral mergers, then the present-day galaxy with its $\sim8000$ old-halo clusters
would have required the merger of $\sim$45 large spirals the size of the 
Milky Way (each with $100 - 200$ GCs) 
{\it if} no new globulars were produced.  This is an uncomfortably 
large requirement.  However, it is more likely that
some GC formation occurred during the mergers.  If we suppose that roughly half 
the GCs were formed out of incoming gas in this manner, 
then we might need $\sim 25$ such spirals, each bringing in 
$\sim 2 \times 10^9 M_{\odot}$ of gas for a reasonable 
GC formation efficiency (see \markcite{har95}Harris 1995). These numbers translate to a rate averaging at least 3 mergers/Gyr. While this is not 
impossible, one must assume a very rich initial population of big spirals and 
a central galaxy adept at capturing them. In addition, recent evidence (\markcite{and94}Andreon 1994) shows that previous cluster surveys have underestimated the number of spirals by a factor two to three. If the Perseus cluster presently contains many spiral galaxies, and the above scenario is correct, the initial population must have been overwhelmingly spiral-rich.

Conversely, if the original central object was a large elliptical containing half the present GC population, then the necessary spiral accretion rate
could be reduced to a less demanding 1/Gyr. Yet to possess an original specific frequency $S_N \sim 15$, proto-NGC 1275 would also be required to be $\geq 4$ times less luminous than it is today. If GC formation during mergers was significant, then the original galaxy must have been even smaller, and alternative (2) then becomes almost indistinguishable from (1). In short, changing a high-$S_N$ galaxy to a low-$S_N$ one puts rather severe demands on any accretion or merger process.  

{\it (3)  NGC 1275 began as a normal-$S_N$ cD or giant elliptical.}  
This scenario is the most conservative of all, since it reduces
the need to invoke mergers at any high rate. Rather than having accretions
dilute or enhance the $S_N$, we suggest that NGC 1275 began as a gE with
a specific frequency similar to that observed today. The mergers required to build up the cD envelope need not have altered the $S_N$ dramatically, as might be the case if other gas-poor gEs of similar $S_N$ were swallowed by the proto-NGC 1275. The present-day activity (nuclear star formation, the HV system, etc.) may then be interpreted as fairly rare occurrences which will have little effect in the long run. The young blue clusters of \markcite{hol92}Holtzman \etal (1992) may dissolve away in the next $\sim 1$ Gyr from tidal shocking, dynamical friction, and self-disruption before they can add to the old-halo cluster population.

Evidence which could verify or refute the above possibilities may be found in the color distribution of the GCS. Unimodality would indicate single-burst formation as in the case of ELS-model formation (\markcite{els62}Eggen, Lynden-Bell \& Sandage 1962), while bimodality could point to multiple-epoch formation. Furthermore, if a large fraction of NGC 1275's GCs came from spirals originally, then they should be metal-poor and thus show up as a significant population of bluer objects than normal for a giant elliptical.
 
None of the alternatives discussed above is entirely satisfactory, since the small old-halo cluster population in NGC 1275 remains unusual in the context of other cD-type galaxies. Although central cDs with normal or
subnormal $S_N$ values ($< 5$) are known elsewhere, none have the
level of central activity that NGC 1275 exhibits (see \markcite{hpm95}HPM95).
For further clues, it is natural to turn to the combination of features associated with the central regions of this object.
It is agreed to be a Seyfert galaxy (\markcite{ver94}Vermeulen \etal 1994, 
\markcite{nes95}Nesterov \etal 1995) with a blazar-like spectrum (\markcite{les95}Lester \etal 1995). 
Its highly active nuclear region contains the radio source 3C 84 which, 
after a burst of activity observed in 1980, 
was the brightest extragalactic source at mm wavelengths. 
\markcite{ver94}Vermeulen \etal (1994) have been observing the nucleus 
in the 22 GHz range since 1981 and have recently discovered a counterjet 
\footnote{A separate phenomenon from that discussed in the Introduction.} 
to complement the previously known relativistic nuclear jet. 
\markcite{ver94}Vermeulen \etal believe that the counterjet eluded discovery for some time 
due to a toroid of gas about the central engine (assumed to be a black hole) 
that obscured the counterjet through free-free absorption.  
Presently, the core luminosity is $\sim$ 10$^{43}$ ergs s$^{-1}$, a value in 
the same range as more distant radio galaxies. 

Recent theoretical modelling (\markcite{har94}Harris \& Pudritz 1994) makes the case that
globular clusters form out of cool gas collected within supergiant
molecular clouds (SGMCs). An obvious speculation is then that if
the nucleus of NGC 1275 was even more active at an earlier epoch,
the high ionizing radiation and radiation pressure could have disrupted the
normal GC formation process. That is, by the time the nuclear activity
died down to the point that star formation could proceed more vigorously,
the host SGMCs might already have broken down into smaller units 
within which the formation of massive star clusters 
was impossible (see \markcite{har94}Harris \& Pudritz).  
However, the problem with invoking this sort of process to inhibit GC formation
is that most (and probably all) of the central giant cD's 
should have central massive black holes (e.g. \markcite{fab95}Fabian \& Rees 1995; \markcite{harm94}Harms \etal 1994;\markcite{for94} Ford \etal 1994).  That is, most or all of them
should have experienced highly active
early phases of disruptive nuclear activity.  Yet some (M87, for example) successfully
produced huge numbers of GCs, while others (such as NGC 1275) could not.
In addition, it is plainly true that massive star clusters have formed quite
recently in the inner few kpc of NGC 1275 {\it despite} its strong nuclear 
activity; thus it is not clear just how these seemingly disruptive
effects actually influence GC formation.

Another piece of the puzzle may lie in the details of the Perseus environment 
and the X-ray halo gas, which provides a signature of the evolutionary state
of the cluster.  Actively studied at X-ray and optical wavelengths, 
A 426 was at first assumed to be relaxed (\markcite{jon84}Jones \& Forman 1984). However, more extensive recent data favor a picture of a more disordered 
cluster which has not yet virialized. 
\markcite{sle94}Slezak, Durret, \& Gerbal (1994), \markcite{moh93}Mohr, Fabricant, \& Geller (1993), and 
\markcite{mcm89}McMillan, Kowalski, \& Ulmer (1989) 
have found evidence of substructure from ROSAT and Einstein studies of 
the cluster. \markcite{and94} Andreon (1994) cites the non-uniform distributions of various Hubble types and along with \markcite{bra81}Branduardi-Raymont \etal (1981), 
the observation that the optical and X-ray centers of A 426 
do not coincide. The analysis of \markcite{sch92}Schwarz \etal (1992) shows a cool subcluster 
in the eastern region of A 426, with an X-ray temperature 
$> 2$ keV less than the mean cluster average of 6-7 keV. 
This feature lies within 1 Mpc of NGC 1275. In other words, even
though NGC 1275 is rather clearly sitting close to the dynamical center
of its very rich surrounding region, that region has just as clearly 
not yet settled into equilibrium.  
If the Perseus cluster itself is an amalgamation of many
subclusters, then we could reasonably suggest that NGC 1275 started out in one
of these smaller clusters, in which lower-$S_N$ ellipticals are quite common
(\markcite{har91}Harris 1991). But as a counterargument,
examination of some prototypical high-S$_N$ galaxies such as M87 in 
Virgo (\markcite{bin89}Binggeli \& Tammann 1989), 
or NGC 4874 in Coma (\markcite{moh93}Mohr \etal 1993) shows that they too belong to
clusters which display plain evidence for substructure.

These arguments lead only to {\it possible} links with galaxy environment and
environmental history.  On the available evidence, we are forced to
conclude that {\it the specific frequency of a cD galaxy 
is not linked to the evolutionary state of its host cluster} in any obvious
way.  In other words, $S_N$ seems more likely to be a result of the
initial conditions of the host gE galaxy.

The most conservative synthesis of this discussion is to propose a scenario in which NGC 1275 originated as a large, or moderately large, E galaxy within a cluster that was significantly smaller than A 426 is today.  There, proto-NGC 1275 would have had a GCS of average specific frequency. Over time, it may have merged with other galaxies, perhaps enough to cause a modest decrease in the number of GCs per unit luminosity. At the same time, it was fortunate enough to become the visible center of the larger Perseus cluster as other galaxy subclusters were amalgamated, building up to the system we now see.  In this scenario, we would interpret the extensive nuclear activity in the center of NGC 1275 as providing an intriguing window on cluster formation processes, but perhaps as inconsequential to the $S_N$ of the system, since the globular clusters formed there will not survive long or add significantly to the old-halo population.  

This general interpretation probably makes the least demands on the
ill-understood details of the merger process (either the merger rate itself
or the cluster formation efficiency, for which we do not yet have a 
detailed physical theory).  However, it is clear that other, more extreme,
interpretations cannot be strongly ruled out.

\section{Summary}

Our deep R-band survey of the old halo of NGC 1275 shows 
that this galaxy has an low-to-average population of globular clusters despite its central location and dominant position within the Perseus cluster.
Various arguments suggest that NGC 1275 might be anything from a ``starpile'' of many previous mergers, to an initially enormous galaxy that has undergone only minor accretions since. In our view, the most likely possibility is that NGC 1275 started out with much the same specific frequency ($S_N \sim 4$) as it has today. Results of this study indicate that GC formation is connected neither with a galaxy cluster's cooling flow, nor the evolutionary state of its cluster environment. 
Unfortunately, we are still left without any firm indications as to the formation mechanisms of high-$S_N$ galaxies.
Additional data for the NGC 1275 GC system in the form of accurate 
photometric color indices for the old-halo clusters would help to clarify
some aspects of the argument.

\acknowledgements

DK would like to thank Jeff Secker, Dean McLaughlin, and Pat Durrell for offering the use of their algorithms and for their enlightening discussions concerning all things globular. She also wishes to thank the referee, Steve Zepf, as well as Ralph Pudritz and Doug Welch for their helpful suggestions. This work was supported in part by the Natural Sciences and Engineering Research Council of Canada, through operating grants to WEH and HBR.


\newpage
\begin{center}
Figure Captions
\end{center}

\figcaption[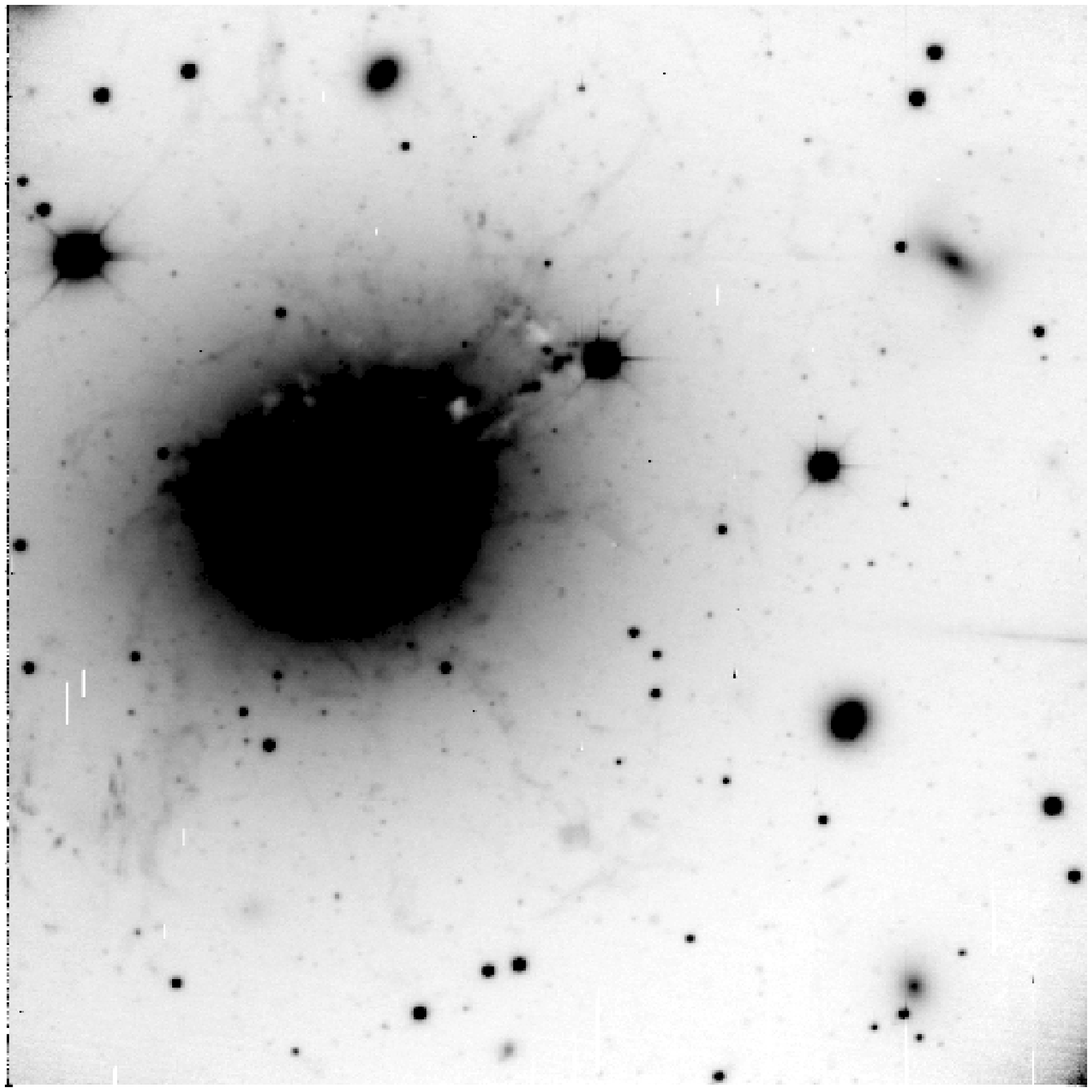]{The field surrounding NGC 1275 as imaged with CFHT and HRCam. The field of view is 2.7$\min$ on a side; this image is a composite of 20 300s exposures in the $R$ band. \label{galaxy}}

\figcaption[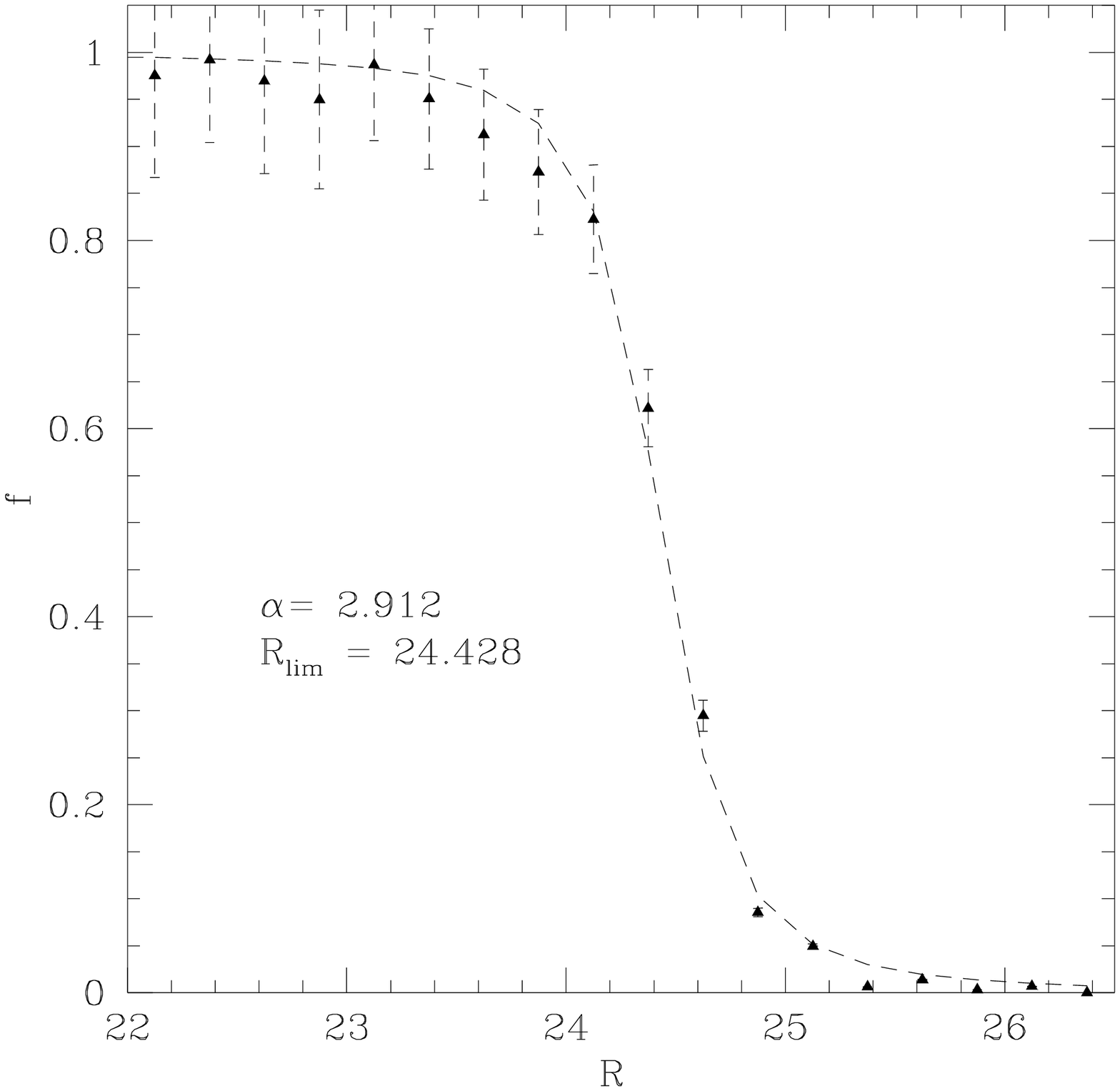]{Completeness of our $R$-band photometry
as a function of magnitude, as derived from artificial-star simulations and
DAOPHOT.  The two-parameter interpolation function (see text) is shown
as the line through the data points.  The limiting magnitude, defined
as the magnitude at which $f=0.5$, is at $R \simeq 24.5$. \label{compfunc}}

\figcaption[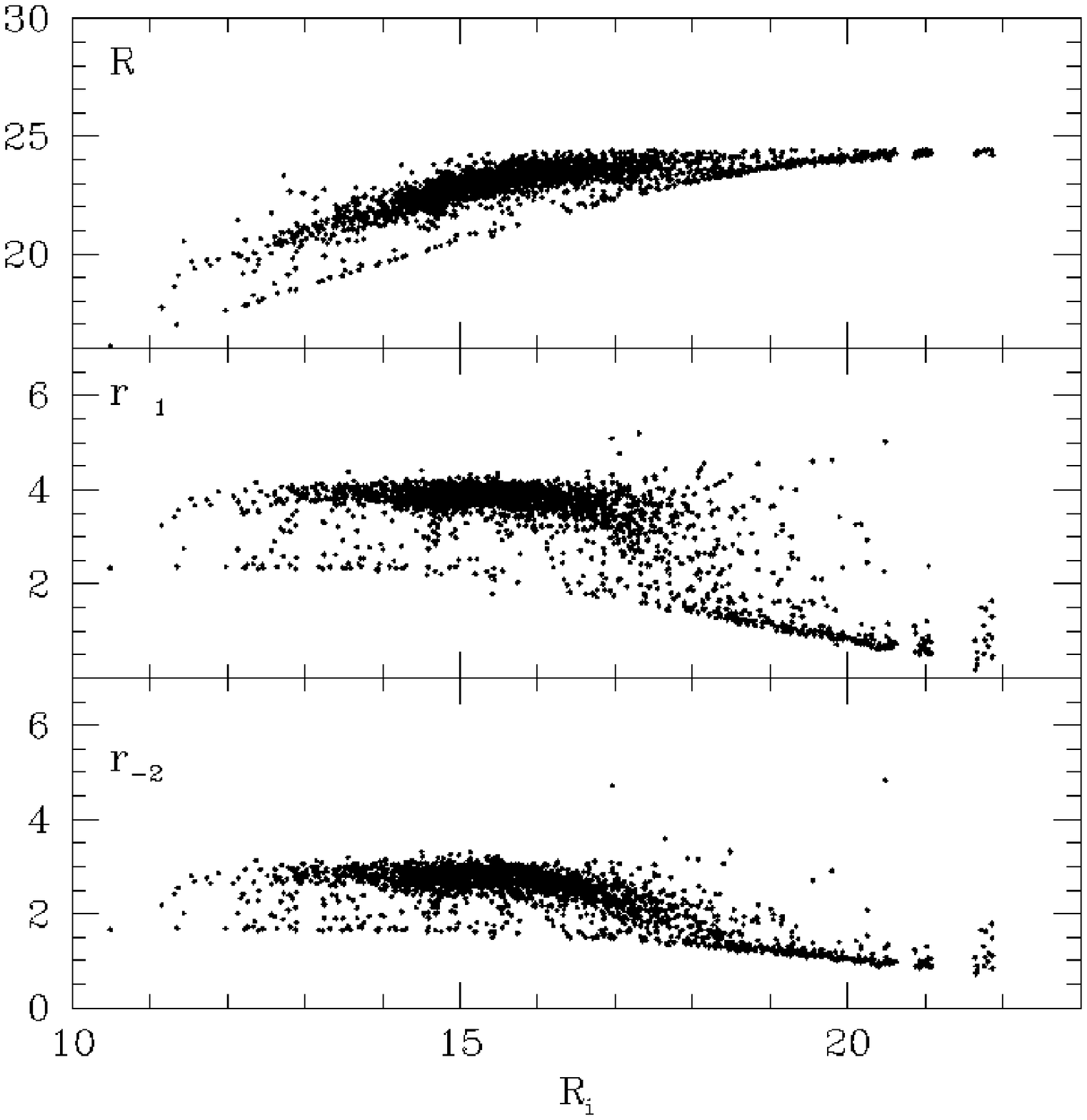]{Image classification moments for the measured objects in the program field.  Here $R_i$ is an isophotal magnitude, $r_1$ a radial moment measuring image profile wingspread, and $r_{-2}$ a radial moment measuring central peakedness (from Harris \etal 1991).  Most detected objects in the frame are visibly nonstellar, with the `starlike' objects showing up as the narrow sequence in the lower part of each panel.  Objects with $r_1 > 2.5$ px are rejected from the sample; see text. \label{moments}}

\figcaption[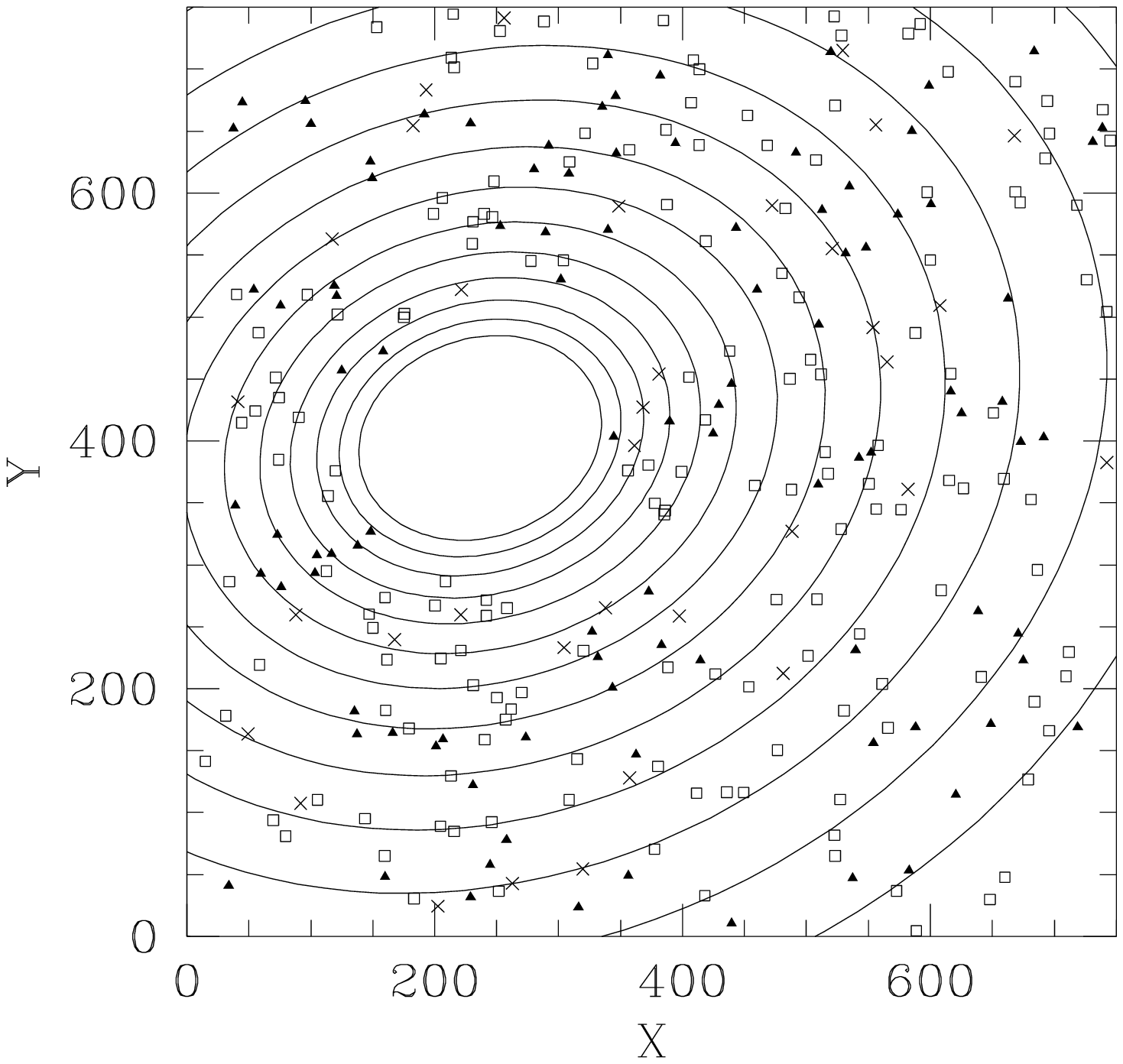]{Spatial locations of globular
cluster candidates (faint, starlike objects) around NGC 1275.  
The elliptical annuli, centered on the galaxy, follow the shape of the galaxy 
light and are used to calculate the radial distribution
shown in the next figure. Objects with $22 < R \leq 23$ are represented 
by crosses, $23 < R \leq 24$ by filled triangles, and $R >$ 24 by squares. \label{xypos}}

\figcaption[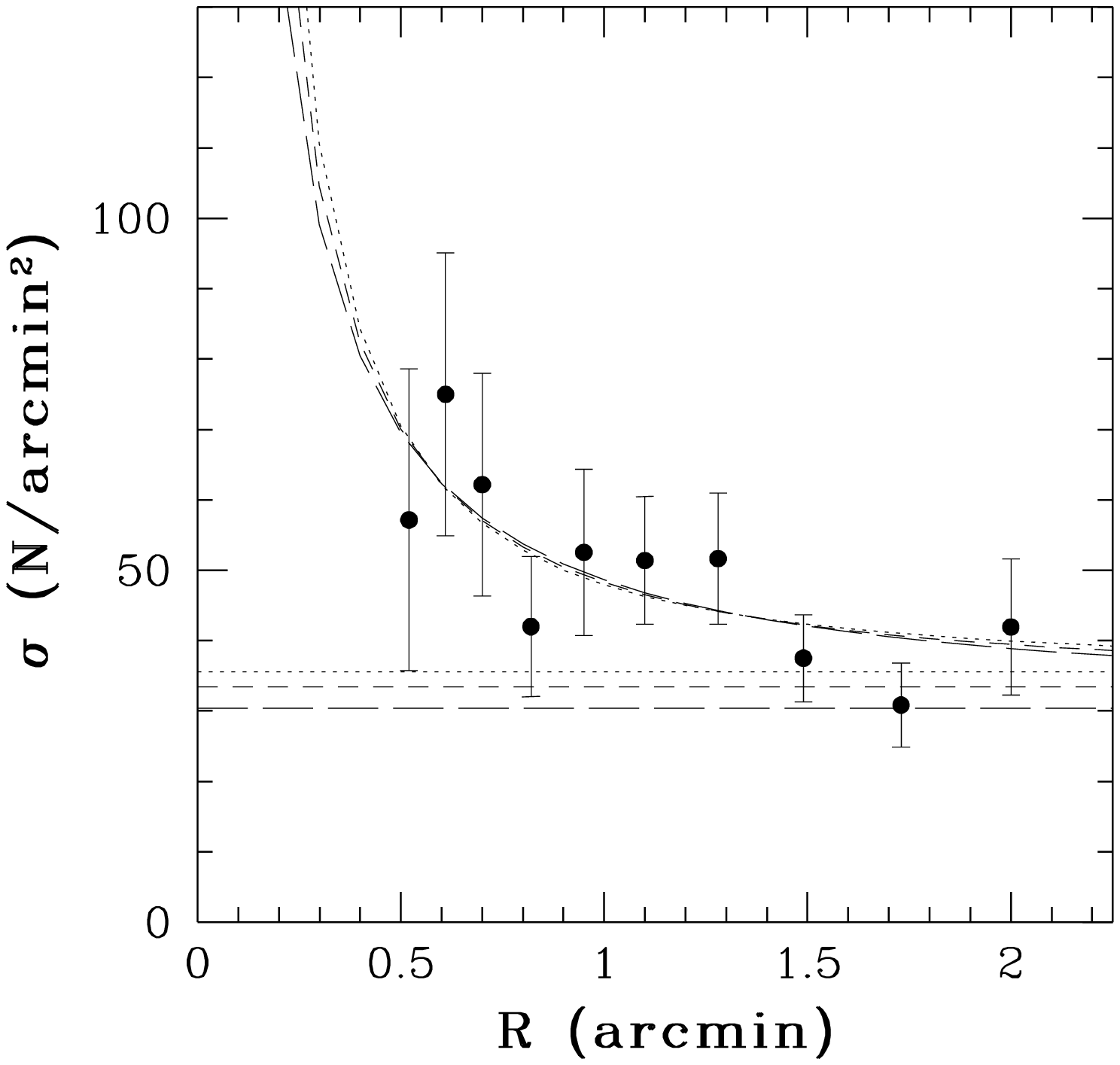]{Radial distribution of globular cluster candidates around NGC 1275.  Here $\sigma$ is the number of starlike images fainter than $R = 22$ per arcmin$^2$, plotted as a function of radius (annular semimajor axis).  The horizontal dotted lines show the adopted level of the background for $\alpha = -1.1$ (dotted line), $\alpha = -1.3$ (short dashes), and $\alpha = -1.5$ (long dashes). The corresponding power law fits are very similar to each other. The innermost two annuli of the previous figure have been excluded from this graph due to poor number statistics. \label{sigma3}}

\figcaption[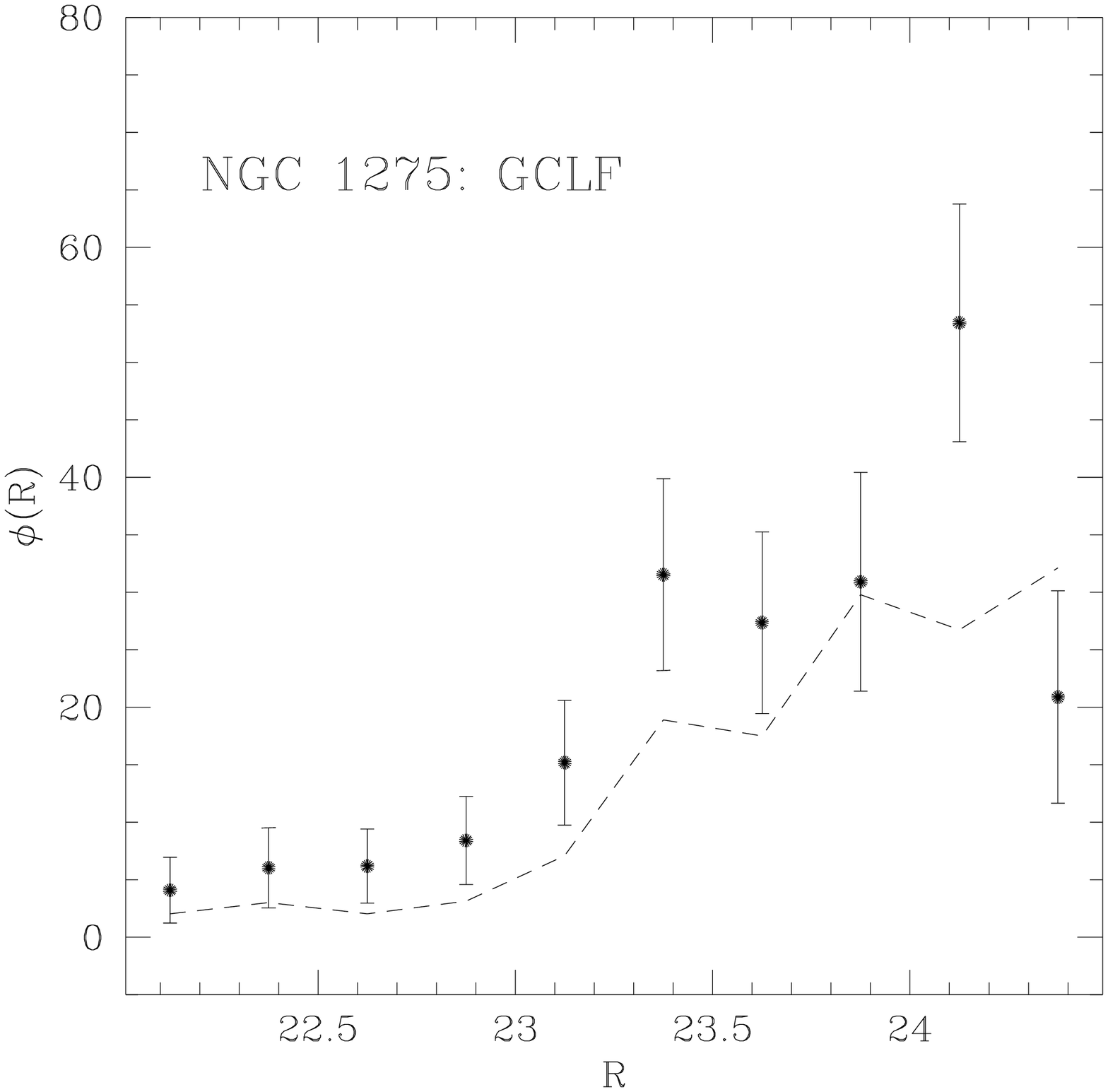]{The luminosity function (GCLF) for the old-halo
globular cluster population in NGC 1275.  The solid symbols
with error bars show the residual GCLF $\phi(R)$ (number per quarter-magnitude
bin) after subtraction of background. The adopted background LF,
defined by the three outermost annuli in Fig.~5,
is shown for comparison by the dashed line. \label{lumfunc}}

\figcaption[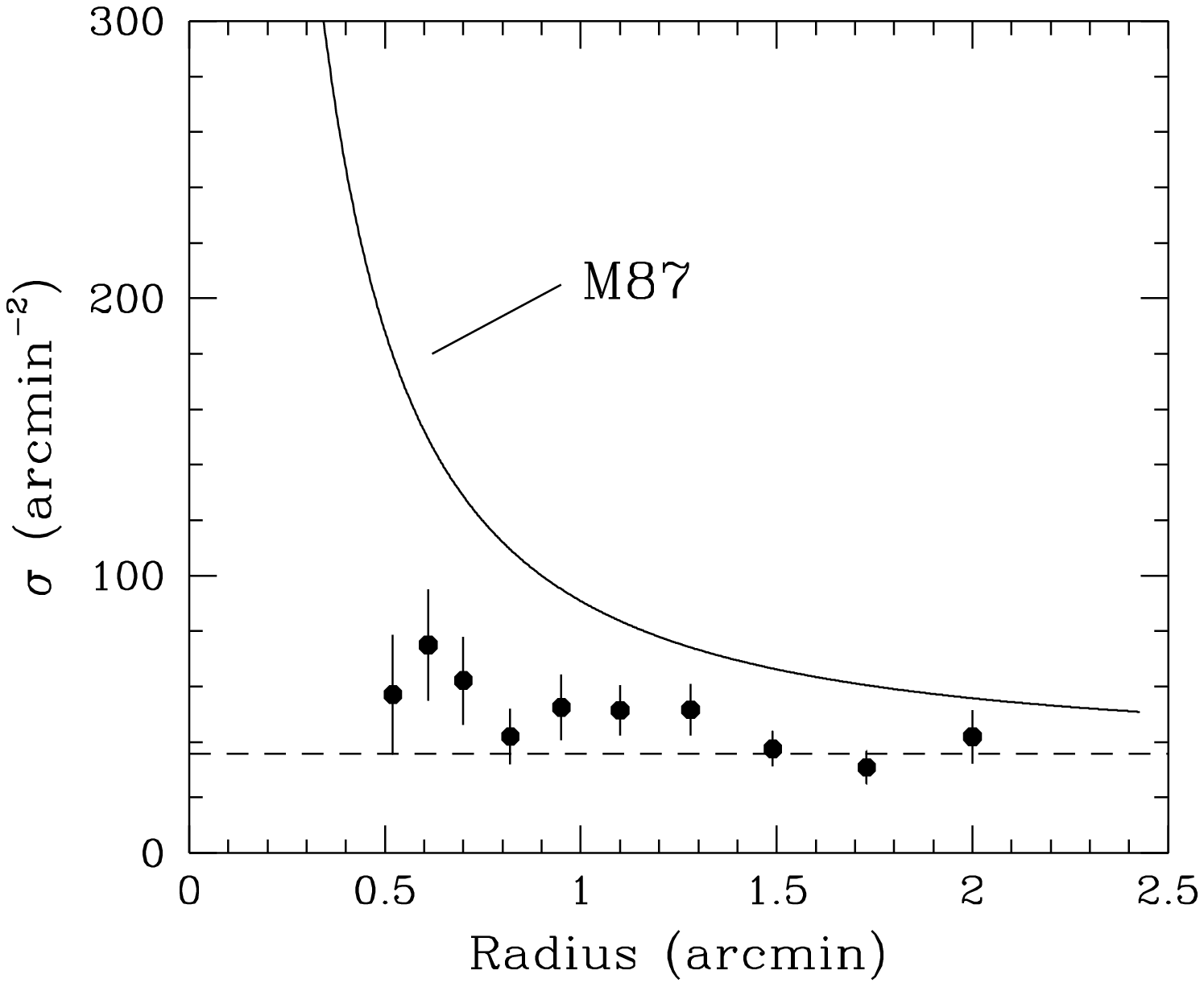]{Radial distribution of the GCS, replotted
for comparison with M87.  The solid dots show the observed points as
in Fig.~6 above.  The labelled line shows the distribution that would have been 
observed if NGC 1275 had the same specific frequency as M87, measured
to the same limiting magnitude (data from McLaughlin \etal 1994). \label{sigmam87}}

\end{document}